\documentclass[pra,showpacs,twocolumn,longbibliography,secnumarabic,superscriptaddress]{revtex4-2}

\usepackage{graphicx}
\usepackage{amssymb}
\usepackage{float}
\usepackage[T1]{fontenc}
\usepackage{amsmath}
\usepackage{mathtools}
\usepackage{url}
\usepackage{hyperref}
\usepackage[dvipsnames]{xcolor}

\usepackage{amssymb}
\usepackage{dcolumn}% Align table columns on decimal point
\usepackage{bm}% bold math
\usepackage{soul}

\usepackage{xfakebold}

\newcommand{\fbseries}{\unskip\setBold\aftergroup\unsetBold\aftergroup\ignorespaces}
\makeatletter
\newcommand{\setBoldness}[1]{\def\fake@bold{#1}}
\makeatother

\begin{document}

\title{Off-resonant Fano–enhanced single-molecule-resolution imaging  with a CW source}% Force line breaks with \\

\author{Rasim Volga Ovali}%
\affiliation{Department of Physics,  Recep Tayyip Erdogan University, 53100 Rize, Turkey}%

\author{Taner Tarik Aytas}%
\affiliation{Department of Physics, Faculty of Science, Akdeniz University, 07058 Antalya,Turkey}%

\author{Ramazan Sahin}%
%\email{rsahin@itu.edu.tr}
\affiliation{Department of Physics, Faculty of Science, Akdeniz University, 07058 Antalya,Turkey}%

\author{Mehmet Emre Tasgin}%
 \email{metasgin@hacettepe.edu.tr}
\affiliation{Institute of Nuclear Sciences, Hacettepe University, 06800 Ankara, Turkey}%
%\affiliation{Department of Nanotechnology and Nanomedicine, Faculty of Science, Hacettepe University, 06800 Ankara, Turkey}%

\date{\today}% It is always \today, today,
             %  but any date may be explicitly specified

\begin{abstract}
Apertureless scanning near-field optical microscopy (a-SNOM) is typically limited to $\sim$10 nm resolution by the tip apex size. We demonstrate that $\sim$1-nm resolution can be achieved under {\it continuous-wave} (CW) illumination by exploiting Fano path interference. A defect center that naturally forms at the apex of a metal-coated AFM tip acts as a quantum object and induces Fano interference, forcing a stronger but normally off-resonant plasmonic mode (597 nm) to operate effectively on resonance at the driving wavelength (520 nm). Because this interference occurs only beneath the defect, a $\sim$1-nm-wide, strongly enhanced near-field hotspot is created. Using this off-resonant Fano-enhanced field, we achieve single-molecule-resolution imaging based on exact three-dimensional Maxwell simulations.
%	Scanning near-field optical microscopy can detect refractive-index changes with a spatial resolution limited by the size of the tip apex, which can be as small as 10 nm for high-quality metal-coated AFM tips. Here, we show that imaging  with $\sim$1-nm resolution can be achieved using a {\it continuous-wave} (CW) source by exploiting Fano path interference. In our setup, the metal-coated AFM tip is covered with a 2D material, and a defect center naturally forms at the sharpest point of the apex. This defect generates a Fano interference effect: it makes a stronger but normally off-resonant mode (597 nm) effectively operate resonantly another (driving, 520 nm) frequency. This takes place only beneath the defect --so the field beneath the defect ($\sim$1 nm) is substantially stronger than the rest of the apex.
%	 By using this off-resonant Fano-enhanced hotspot located directly under the defect, we achieve single-molecule-level imaging. All results are obtained from exact solutions of the 3D Maxwell equations.
\end{abstract}

%Use showkeys class option if keyword
                              %display desired
%\keywords{Suggested keywords}%Use showkeys class option if keyword
                              %display desired
\maketitle

%\tableofcontents

%%%%%%%%%%%%%%%%%%%%%%%%%%%%%%%%%%%%%%%%%%%%%%%%%%%%%%%%%%%%%%%%%%%%%%%%%%%%%%%%%%%%%%%%%%%%%%%%%%%%%%%%%%%%%%%%%%%%%%%%%%%%%%%%%%%%
%\section{Introduction}

Metal nanostructures (MNSs) can localize incident electromagnetic fields into nanometer-sized hotspots, where light–matter interaction is enhanced by several orders of magnitude~\cite{hoppener2012self}. This effect is widely used, for example, in thin gold film–based biosensors~\cite{Deka2018,dunn1999near,yanik2011seeing}, which are common in medical laboratories today. In such devices, antibody-bound viruses modify the refractive index near the gold film, and this change is detected through a shift in the reflection angle~\cite{Homola2008,altug2022advances}.

The same phenomenon can also be used for subwavelength imaging. A gold-coated atomic force microscope (AFM) tip can localize the incident field at its apex~\cite{hillenbrand2000complex}, which can be as small as 10 nm~\cite{alpan1,bazylewski2017review}. Scanning a surface with such a tip not only enables optical imaging with 10-nm resolution~\cite{alpan1,bazylewski2017review} but also provides refractive index information at that scale~\cite{hillenbrand2000complex} and can be used in subwavelength lithography~\cite{long20233d,zeng2017deep}. This technique has even enabled the experimental demonstration intriguing physics such as negative phase velocity at 2D hexagonal boron nitrides~\cite{yoxall2015direct,liu2016nanoscale}.

Although this 10-nm resolution is well below the diffraction limit of optical microscopes (half the wavelength), two main challenges remain:
(i) achieving such resolution requires advanced fabrication facilities, so in most laboratories, the apex size is typically around 50 nm; and
(ii) it is still natural to ask whether optical (near-field) imaging with even finer resolution can be achieved.

\begin{figure}
	\centering
	\includegraphics[width=\linewidth]{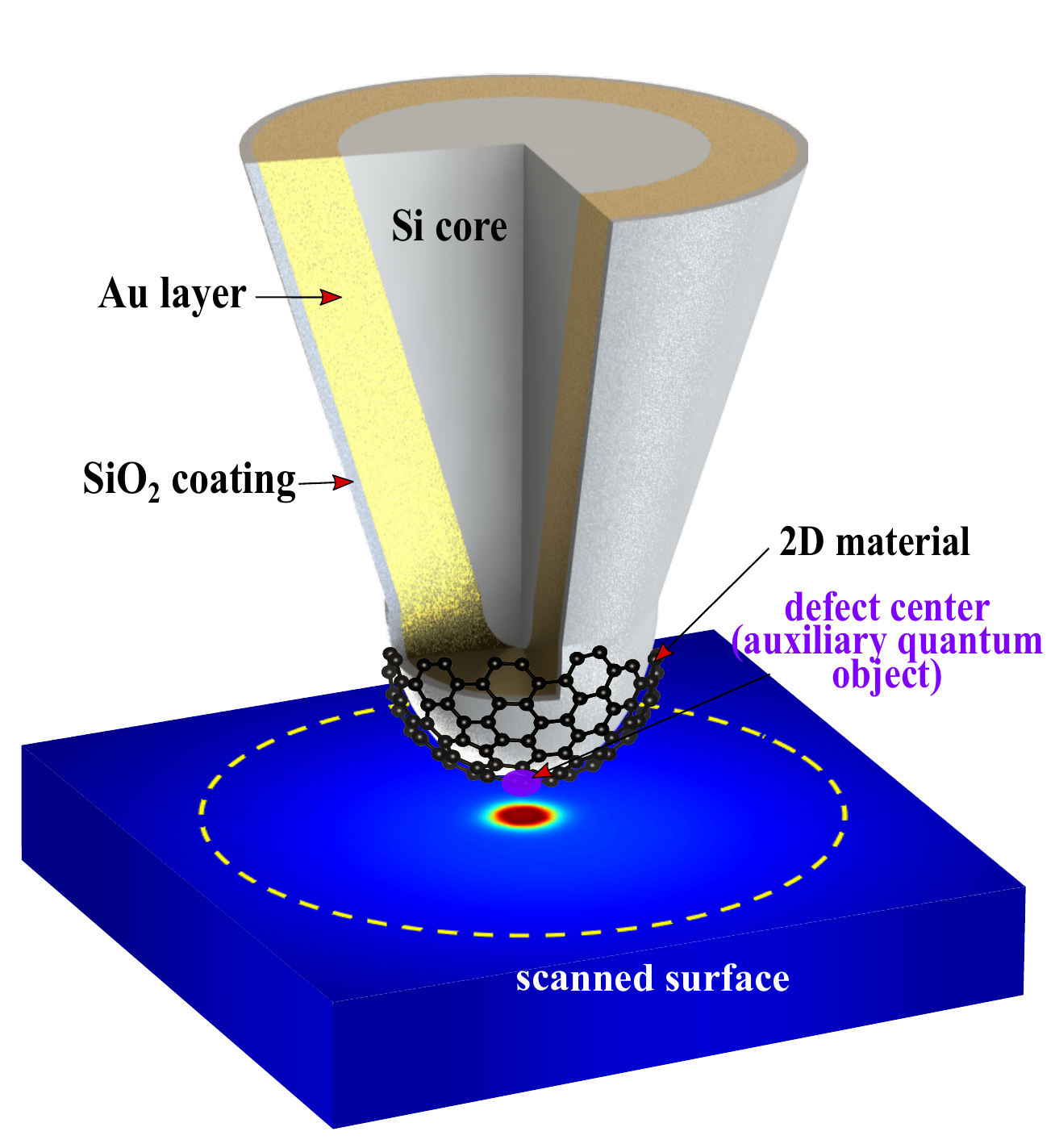}
	\caption{A gold-coated AFM tip is covered with a 2D material~\cite{palacios2017large}. A stress-induced defect center naturally forms in the 2D layer at the point of maximum bending—the tip apex. This defect center, which acts as a quantum object (QO), possesses strong oscillator strength and behaves as a single emitter~\cite{branny2017deterministic}. The narrow-linewidth QO located at the apex hotspot introduces a Fano interference channel localized beneath it. When the tip is illuminated by a 520 nm continuous-wave (CW) source, this Fano enhancement—derived analytically in the text—forces a much stronger plasmonic resonance that normally appears at 597 nm to operate effectively at the off-resonant wavelength of 520 nm (see Fig.~\ref{fig2}b). Because this interference pathway exists only at and directly beneath the QO, the near-field intensity below the defect becomes substantially stronger than at the rest of the apex. As a result, a $\sim$1-nm-wide Fano-enhanced intensity peak forms (Fig.~\ref{fig3}c) and scans across the sample surface, enabling $\sim$1-nm-resolution a-SNOM imaging (Fig.~\ref{fig4}). The intensity maps shown are exact solutions of the three-dimensional Maxwell equations obtained using Lumerical simulations~\cite{Ansys2024}. The dashed yellow circle denotes the hotspot size of the bare tip without a QO at the apex, shown for reference in Fig.~\ref{fig3}a.
		%A gold-coated AFM tip is covered with a 2D material~\cite{palacios2017large}. A stress-induced defect center forms in the 2D layer at the point of maximum bending—the tip apex. This defect center (the quantum object, QO) possesses strong oscillator strength and behaves as a single emitter~\cite{branny2017deterministic}. The narrow-linewidth QO located at the apex hotspot introduces a Fano interference channel beneath it. When the tip is illuminated with a 520 nm CW source, this Fano enhancement—derived analytically in the text—causes the stronger resonance that normally appears at 597 nm to operate effectively at the off-resonant wavelength 520 nm (see Fig.~\ref{fig2}b). Because this interference pathway occurs only at and beneath the QO, the field directly under the defect becomes substantially stronger than the rest of the apex. This produces a $\sim$1-nm-wide Fano-enhanced intensity peak (Fig.~\ref{fig3}c) that scans across the sample surface that achieves a $\sim$1-nm-resolution aSNOM (Fig.~\ref{fig4}). The intensity maps shown are exact solutions of the 3D Maxwell equations obtained using Lumerical simulations~\cite{Ansys2024}. The dashed black circle denotes the hotspot size of the bare tip without a QO at the apex; this reference hotspot is shown in Fig.~\ref{fig3}a.
	}
	\label{fig1}
\end{figure}

In a recent work~\cite{ovali2021single}, we proposed a method that reduces the effective hotspot size down to the scale of a single molecule. This method relies on plasmon lifetime enhancement~\cite{stockman2010dark,yildiz2020plasmon,asif2024voltage,ovali2025surface}, which occurs around a quantum object (QO) placed at the apex of a gold-coated tip. Using a pulsed excitation source, the near-field below the QO persists for much longer times, allowing signal collection from a $\sim$1-nm region. The QO is a stress-induced defect center in a 2D material (see Fig.~\ref{fig1}). The gold-coated apex is covered by the 2D material~\cite{palacios2017large,branny2017deterministic}, which bends slightly due to the coating. The defect center forms at the sharpest point of the apex—its lowest position, where the bending is maximum (see Fig.~\ref{fig1}).

However, using ultrashort pulses complicates practical imaging applications. The effective hotspot intensity varies in time, and the near-field below the quantum object (QO) does not fully decay before the next pulse arrives. This can cause unwanted interference between consecutive pulses. Therefore, developing a similar method that works with continuous-wave (CW) lasers~\cite{alpan1}—which is the conventional approach in imaging—would represent significant progress for practical applications.

\begin{figure}
	\centering
	\includegraphics[trim={0 1cm 0 0.5cm}, clip, width=\linewidth]{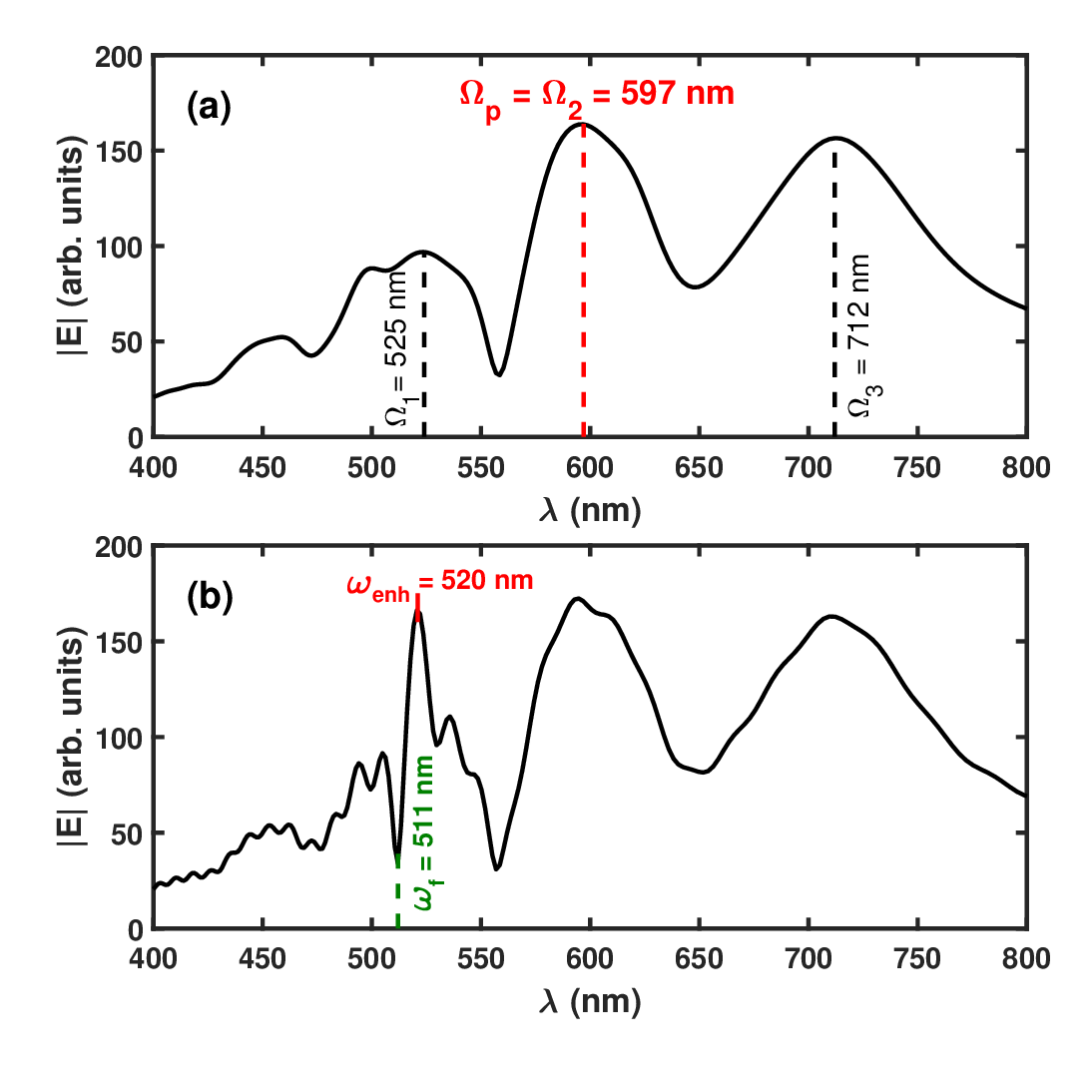}
	\caption{ Near-field spectrum just below the apex in the (a) absence and (b) presence of the quantum object (QO).
		(a) The gold-coated tip exhibits three resonances.
		(b) When a QO with level spacing $\Omega_{\rm \scriptscriptstyle QO} = 525$ nm is placed beneath the apex, the near-field spectrum shows two Fano interference effects: (i) a retardation-shifted Fano transparency at $\omega_{\rm f} = 511$ nm and (ii) a {\it Fano enhancement} at $\omega_{\rm enh} = 520$ nm, which we use for off-resonant a-SNOM imaging in Fig.~\ref{fig4}. Notably, the near-field response at $\omega_{\rm enh} = 520$ nm is as strong as the response at $\Omega_p\equiv\Omega_2=$597 nm. This occurs because the second term in the denominator of Eq.~(\ref{FR}) cancels the off-resonant contribution $i(\Omega_p - \omega)$, effectively bringing the stronger $\Omega_p\equiv\Omega_2=$597-nm peak into resonance at $\omega_{\rm enh} = 520$ nm.
		When the tip is illuminated at $\omega = \omega_{\rm enh} = 520$ nm, a $\sim$1-nm-wide {\it Fano-enhancement spot} appears in the hotspot profile (Fig.~\ref{fig3}c). The former effect, the Fano transparency at $\omega_{\rm f} = 511$ nm, produces a $\sim$1-nm-wide intensity gap (Fig.~\ref{fig3}b).
	}
	\label{fig2}
\end{figure}

In this paper, we propose an alternative and previously unexplored approach that relies on Fano enhancement at a normally off-resonant frequency~\cite{tacsgin2018fano,singh2016enhancement,luk2010fano,adato2013engineered,turkpence2014engineering,postaci2018silent} induced by Fano path interference~\cite{luk2010fano,leng2018strong}. A quantum object (QO)—more precisely, the Fano interference it creates—can make a much stronger intensity mode (the main resonance at 597 nm in Fig.~\ref{fig2}a) behave as if it were resonant at a very different wavelength (520 nm in Fig.~\ref{fig2}b). As a result, when the AFM tip is illuminated at 520 nm, the near-field directly beneath the QO acts as strongly as the 597-nm peak, while the rest of the apex responds with the weaker 520-nm intensity shown in Fig.~\ref{fig2}a.

\begin{figure}
	\centering
	\includegraphics[trim={0 0.5cm 0 0.5cm}, clip, width=\linewidth]{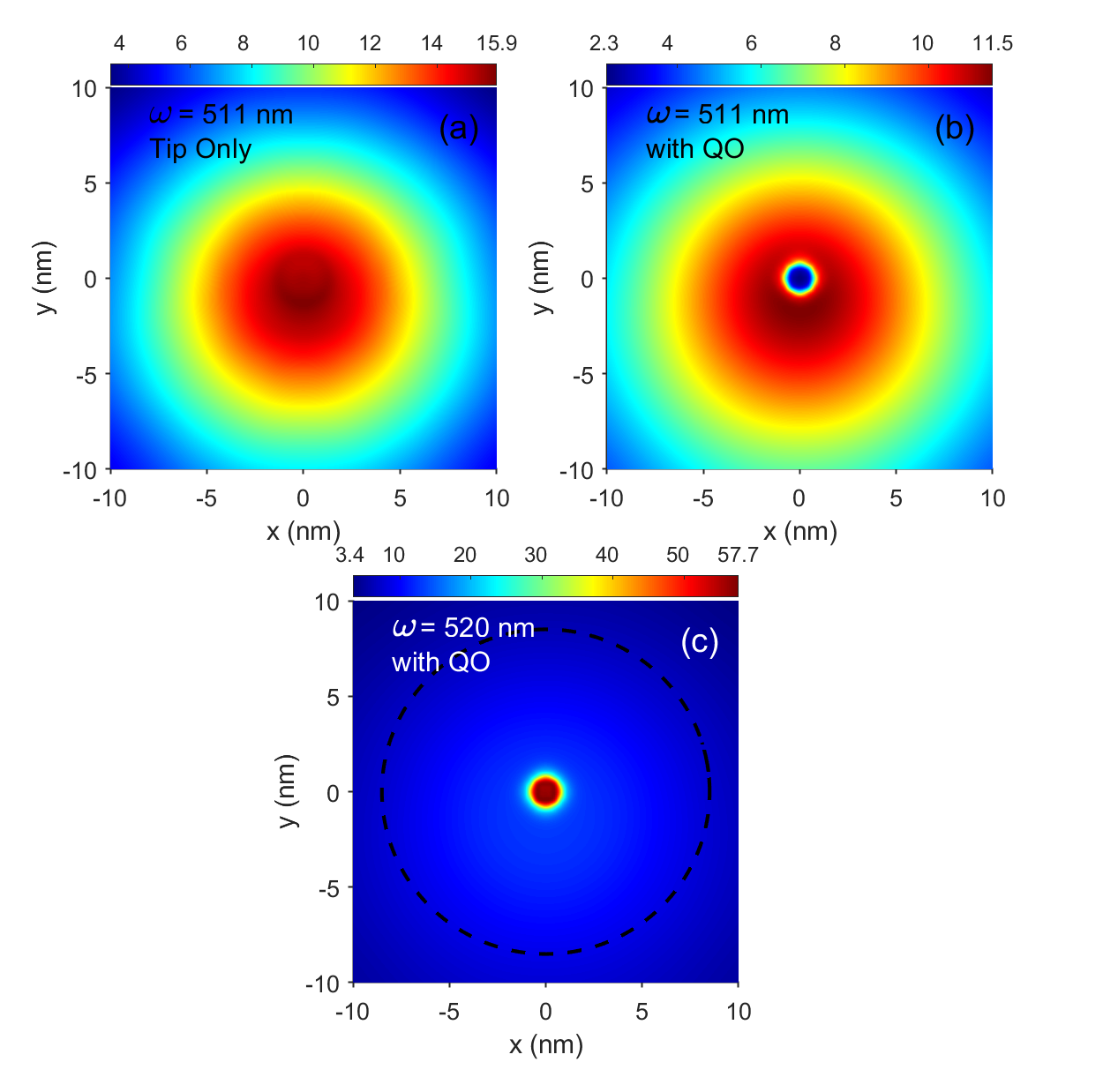}
	\caption{Intensity profile beneath the tip apex in the (a) absence and (b, c) presence of the QO.
		(b) When the tip is illuminated at the Fano-transparency frequency $\omega = \omega_{\rm f} = 511$ nm (as identified in Fig.~\ref{fig2}b), a $\sim$1-nm-wide intensity gap appears inside the hotspot. The intensity within this gap is only about 4\% of the surrounding hotspot field.
		(c) In contrast, when illuminated at the Fano-enhancement frequency $\omega = \omega_{\rm enh} = 520$ nm, a $\sim$1-nm-wide Fano-enhanced spot forms inside the hotspot. This enhancement originates from a path-interference effect that makes the stronger 597 nm resonance effectively operate at 520 nm. The enhanced spot is approximately six times brighter than the nearby hotspot intensity. This effect is the one utilized to achieve single-molecule–resolution a-SNOM in Fig.~\ref{fig4}. %The plots show the electric-field amplitudes $|E|$.
	}
	\label{fig3}
\end{figure}

\begin{figure}
	\centering
	\includegraphics[trim={0 1.5cm 0 0}, clip, width=\linewidth]{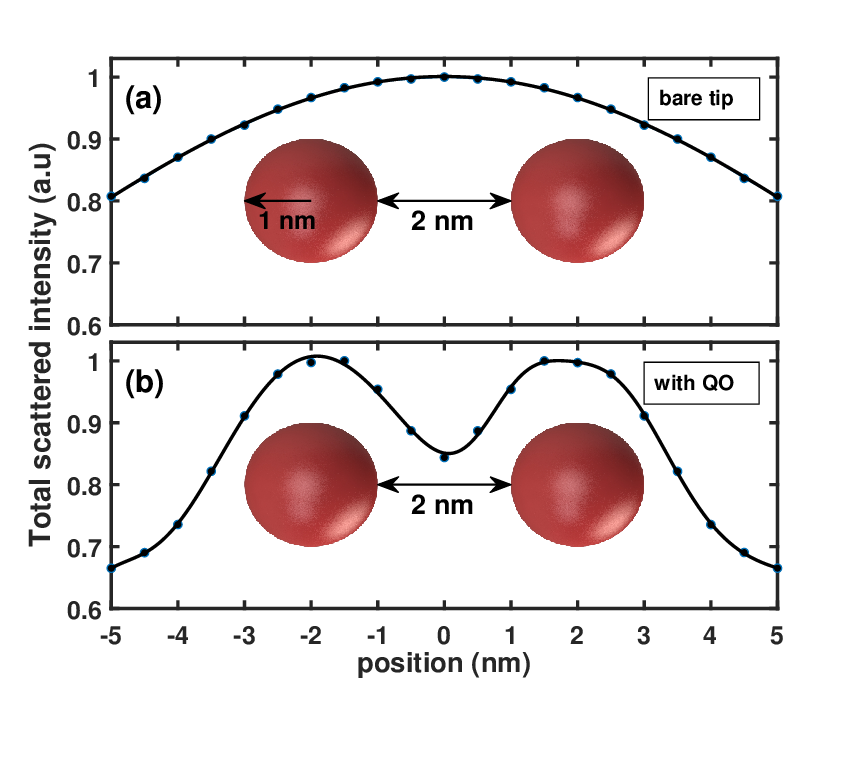}
	% Answer: [trim={left bottom right top},clip]
	%[trim={5cm 0 0 0},clip]
	\caption{The tip shown in Fig.~\ref{fig1} scans two fluorescent nanoparticles (red spheres) in the (a) absence and (b) presence of a QO at the apex under CW illumination at $\omega_{\rm enh}=520$ nm. When the QO is present, the near-field intensity is Fano-enhanced only directly beneath it (see Fig.~\ref{fig3}c), and this localized region dominates the detected signal. As a result, the effective spatial resolution of the scan is significantly improved.  		}
	\label{fig4}
\end{figure}

Under 520-nm illumination, the near-field intensity directly beneath the 1-nm-sized QO is enhanced by up to a factor of six compared to the bare gold-coated tip (Fig.~\ref{fig3}c). We use this effect to achieve single-molecule–resolution apertureless scanning near-field optical microscopy (a-SNOM) with a continuous-wave (CW) source, as shown in Fig.~\ref{fig4}. In addition, through both supplementary simulations and an analytical model, we explicitly verify that this off-resonant Fano enhancement mechanism is indeed responsible for the single-molecule-resolution a-SNOM imaging observed in our results.

A quantum object (QO) located at the tip apex introduces two distinct Fano effects:
(i) a well-known transparency at $\lambda = 511$ nm~\cite{luk2010fano,leng2018strong}, and
(ii) a less-familiar \textit{enhancement} effect at $\lambda = 520$ nm~\cite{tacsgin2018fano,singh2016enhancement,luk2010fano,adato2013engineered,turkpence2014engineering,postaci2018silent,demirtas2025qupers} --the one we employ here.
These features appear in Fig.~\ref{fig2}b, which plots the electric field amplitude $|E|$ measured just below the QO. (i) Under illumination with a $\lambda = 511$ nm CW source, the field amplitude on the plane beneath the tip exhibits a $\sim$1-nm-wide region whose intensity is reduced by a factor of 0.04 (the dark-blue spot in Fig.~\ref{fig3}b). This reduction arises from the Fano-induced transparency.

In contrast, our method exploits the second effect—(ii) the \textit{Fano enhancement} at 520 nm—which is more suitable for a-SNOM imaging. When the same tip is illuminated with a $\lambda = 520$ nm CW source, the enhancement peak in Fig.~\ref{fig2}b generates a $\sim$1-nm-wide intensity spot at the apex --beneath the QO-- that is approximately six times stronger than the usual hotspot intensity (Fig.~\ref{fig3}c). This enhancement occurs because the QO-mediated Fano interference makes the 597 nm resonance (Fig.~\ref{fig2}a) contribute at the off-resonant wavelength 520 nm in Fig.~\ref{fig2}b. Since this off-resonant Fano channel is active only at and beneath the QO, the resulting field spot directly under the QO becomes significantly stronger than the surrounding apex region. Consequently, the ultimate spatial resolution is limited by the size of the defect center itself.

The paper is organized as follows. We first describe the tip geometry shown in Fig.~\ref{fig1}. We then explain how (i) Fano transparency and (ii) Fano enhancement arise within a unified theoretical framework [Eq.~(\ref{FR})] for a general system. The second effect activates the 597 nm plasmonic resonance to operate at 520 nm as if it were on resonance. Next, we present our exact numerical simulations and analyze the signatures of both effects in the near-field spectra (Fig.~\ref{fig2}b). We then examine the near-field intensity distribution and demonstrate both the transparency-induced gap (Fig.~\ref{fig3}b) and the enhancement-induced peak (Fig.~\ref{fig3}c). We further show that the off-resonant Fano enhancement enables 1-nm-resolution a-SNOM imaging (Fig.~\ref{fig4}). Finally, we reexamine the off-resonant Fano enhancement in detail using the parameters of our system and, through supplementary calculations, demonstrate that the 520 nm peak in Fig.~\ref{fig2}b indeed originates from path interference with the stronger 597 nm mode. We conclude by summarizing our findings.

%%%%%%%%%%%%%%%%%%%%%%%%%%%%%%%%%%%%%%%%%%%%%%%%%%%%%%%%%%%%%%%%%%%%%%%%%%%%%%%%%%%%%%%%%%%%%%%%%%%%%%%%%%%%%%%%%%%%%%%%%%%%%%%%%
{\it Tip geometry}.— In a recent study, Atatüre and colleagues~\cite{palacios2017large,branny2017deterministic} demonstrated a method for producing an almost deterministic lattice of single-photon emitters. They deposited tungsten diselenide (WSe${}_2$) and tungsten disulfide (WS${}_2$) monolayers (2D materials) on silicon nanopillars. Stress-induced defects form at the tops of the pillars, where the bending (and thus stress) is maximum. Quantum-confined excitons at these defect sites act like single atoms and exhibit single-photon emission. Density functional theory (DFT) calculations~\cite{ayari2018radiative} show that such defects may possess large oscillator strength and strong polarization response.

In this work, we propose to use this method to obtain a stress-induced defect center at the apex of an AFM tip. A gold-coated silicon AFM tip localizes the incident electromagnetic field at its apex, creating a highly confined hotspot. Our goal is to position a quantum object (QO) at or just beneath the apex so that it strongly interacts with the plasmonic near-field excitation, introducing a Fano resonance~\cite{tacsgin2018fano,wu2010quantum,leng2018strong,shah2013ultrafast,karademir2014plexcitonic}. However, positioning, for example, a molecule precisely at this location is experimentally challenging, and it is difficult to ensure that it remains there during the scanning process.

To overcome this, we propose a method that guarantees the QO remains fixed at the tip apex while scanning. Specifically, we adopt the approach described in Refs.~\cite{palacios2017large,branny2017deterministic}, in which the nanopillars are covered with a 2D material. This configuration naturally forms a stress-induced defect (the QO) at the apex, where the bending is maximum, ensuring it stays in place during scanning. An important advantage of this method is that it does not require the fabrication of ultra-sharp (e.g., 10-nm) gold-coated tips; it remains effective even for larger apex sizes.

A {\it Fano resonance} (FR) arises when a narrow-linewidth quantum object (QO) interacts with a broadband plasmonic excitation~\cite{wu2010quantum,shah2013ultrafast,tacsgin2018fano}. For example, a QO such as a molecule~\cite{leng2018strong} or a defect center~\cite{palacios2017large,BendingDefectAPL2019,darlington2020imaging} placed at the hotspot of a metallic nanostructure (MNS) can induce a transparency window in the plasmonic spectrum. This plasmonic FR is analogous to electromagnetically induced transparency (EIT) observed in atomic ensembles~\cite{fleischhauer2005electromagnetically}.

The QO–plasmon coupling creates two alternative pathways for the absorption and emission of the incident light. In an FR, the energy levels of these two weakly hybridized pathways lie within the plasmon linewidth, so they do not form distinct hybrid modes as in the strong-coupling regime. Because the two pathways oscillate out of phase, their interference cancels absorption and produces a transparency dip in the plasmonic response~\cite{alzar2001classical}.

Fano resonances give rise to three observable effects: one in the time domain and two in steady state.
($\mathfrak{1}$) In the time domain, when a plasmon–QO system is excited by an ultrashort pulse, the plasmon lifetime is extended by the presence of the QO~\cite{stockman2010dark,yildiz2020plasmon,asif2024voltage,ovali2025surface}.
($\mathfrak{2}$) In the steady state, when illuminated by continuous-wave (CW) light tuned to the QO level spacing, $\omega = \omega_{\rm f} \simeq \Omega_{\scriptscriptstyle \rm QO}$, the system becomes transparent.
($\mathfrak{3}$) Interference can also make a natural plasmonic mode behave as if it is resonant at an off-resonant frequency, e.g., making the natural 597 nm mode operate at 520 nm. This is the effect we use in this work.

These steady-state effects—($\mathfrak{2}$), corresponding to (i) in the introduction, and ($\mathfrak{3}$), corresponding to (ii)—can be described by a single analytical expression~\cite{tacsgin2018fano}
\begin{equation}
	\alpha_p=\frac{\varepsilon_p }{[i(\Omega_p-\omega)+\gamma_p] - \frac{y|f|^2}{ i(\Omega_{\rm \scriptscriptstyle QO}-\omega) + \gamma_{\rm \scriptscriptstyle QO} } },
	\label{FR}
\end{equation}
which gives the steady-state plasmon amplitude $\alpha_p$. Here, $\varepsilon_p$ is proportional to the incident pump amplitude; $\Omega_p$ and $\gamma_p$ are the plasmon resonance frequency and damping rate; $\Omega_{\rm \scriptscriptstyle QO}$ and $\gamma_{\rm \scriptscriptstyle QO}$ are the QO level spacing and decay rate; $f$ is the plasmon–QO coupling strength; and $y$ is the QO population inversion. In scaled units (normalized to $\omega$), ``typical'' values are $\Omega_p \sim \Omega_{\rm \scriptscriptstyle QO} \sim 1$, $\gamma_p \sim 0.1$, $\gamma_{\rm \scriptscriptstyle QO} \sim 10^{-5}$, and $f \sim 0.02$–0.1, depending on the oscillator strength and the MNS–QO distance~\cite{wu2010quantum,leng2018strong,singh2016enhancement}.

(i) The well-known Fano transparency occurs at $\omega = \omega_{\rm f} = \Omega_{\rm \scriptscriptstyle QO}$. At this frequency, the second term in the denominator becomes $y|f|^2/\gamma_{\rm \scriptscriptstyle QO}$, which is extremely large because it scales as $1/\gamma_{\rm \scriptscriptstyle QO} \sim 10^5$. This large term dominates the denominator and suppresses plasmon excitation, producing transparency.

(ii) At another illumination frequency, $\omega=\omega_{\rm enh}$, the imaginary part of the second term can cancel the off-resonant contribution $i(\Omega_p - \omega)$. As a result, the plasmon behaves as if it is resonant at $\omega=\omega_{\rm enh}=520$ nm, even though its natural resonance is at $\Omega_p=597$ nm. This yields the lesser-known {\it Fano enhancement}~\cite{tacsgin2018fano} beneath the QO—an intensity about six times stronger than in other parts of the tip apex (Fig.~\ref{fig3}c).
We make use of this effect in this work.

We note that Eq.~(\ref{FR}) is derived from an analytical point-dipole model and does not account for retardation effects. In other words, it assumes that all interactions occur at a single spatial point, neglecting the finite spatial extent of the plasmonic field. Retardation, however, is fully captured in our three-dimensional (3D) finite-difference time-domain (FDTD) simulations, which show that these effects shift the FR frequency $\omega_{\rm f}$ relative to $\Omega_{\rm \scriptscriptstyle QO}$. Both effects (i) and (ii) are clearly observed in our full 3D simulations (Fig.~\ref{fig2}b). We employ the latter effect to obtain a $\sim$1-nm-sized enhanced spot at the tip apex.

Below, we present the exact solutions of the 3D Maxwell equations for the tip configuration presented in Fig.~\ref{fig1}. Then, we explain the Fano enhancement effect using Eq.~(\ref{FR}) and the tip resonances. Giving some spoiler: off-resonant Fano enhancement (cancellations in the denominator of Eq.~(\ref{FR})) carries a stronger intensity mode, of natural resonance $\Omega_p=\Omega_2=597$ nm, to work at resonance at $\omega_{\rm enh}=$520 nm.

{\it Exact solutions of the 3D Maxwell equations}.—We perform full-vector electromagnetic simulations using the finite-difference time-domain (FDTD) package of Lumerical~\cite{Ansys2024}. The simulated geometry is shown schematically in Fig.~\ref{fig1}. Experimental dielectric functions are used for gold and silicon, while the quantum object (QO) is modeled as a Lorentzian emitter with dielectric response~\cite{wu2010quantum,shah2013ultrafast,leng2018strong,LorentzianTMD_Eps2018}
\begin{equation}
	\epsilon_{\rm \scriptscriptstyle L} = \epsilon_b + f_{\rm osc}\frac{\Omega_{\rm \scriptscriptstyle QO}^2}{\omega^2 - \Omega_{\rm \scriptscriptstyle QO}^2 + i \, \gamma_{\rm \scriptscriptstyle QO}\,\omega},
	\label{Lorentzian}
\end{equation}
where $\epsilon_b$ and $f_{\rm osc}$ are the background permittivity and oscillator strength, respectively.
The inner silicon apex has a width of 10 nm, and the outer gold coating has a thickness of 10 nm.

We first compute the electric-field spectrum at a point located just below the apex in the absence of the QO.
Figure~\ref{fig2}a shows three distinct plasmonic resonances supported by the gold-coated tip. 
Next, we place a 1-nm-sized QO directly beneath the apex center and assign it the Lorentzian response in Eq.~(\ref{Lorentzian}), with parameters
$\Omega_{\rm \scriptscriptstyle QO} = 525$ nm,
$f_{\rm osc} = 0.1$,
$\gamma_{\rm \scriptscriptstyle QO} = 10^{10}$ Hz, and
$\epsilon_b = 2.5$
following Refs.~\cite{wu2010quantum,leng2018strong,singh2016enhancement}.
The resulting spectrum just below the QO is plotted in Fig.~\ref{fig2}b.

Two key signatures of Fano interference—corresponding to effects (i) and (ii) discussed beneath Eq.~(\ref{FR})—are clearly visible.
(i) Fano transparency: At $\lambda = 511$ nm, the field amplitude is strongly suppressed, indicating destructive interference between the plasmonic and QO pathways.
(ii) Fano enhancement: At $\lambda = 520$ nm, the interference instead enhances the near field below the QO by bringing the off-resonant stronger plasmon mode at 597 nm into effective resonance at 520 nm.

Figure~\ref{fig3} shows steady-state electric-field amplitude maps in the $x$–$y$ plane below the tip apex.
In Fig.~\ref{fig3}a, a conventional a-SNOM tip illuminated at $\omega = 511$ nm produces a typical diffraction-limited hotspot determined by the apex geometry.
In Fig.~\ref{fig3}b, with the 2D coating and stress-induced QO included, illumination at $\omega = 511$ nm produces a pronounced $\sim$1-nm-wide intensity gap. The field amplitude at the center drops to only $\sim$4\% of the nearby hotspot intensity.
This “dark spot’’ directly reflects the Fano-transparency effect.
(As noted earlier, the observed transparency wavelength is shifted to $\omega_{\rm f} = 511$ nm due to retardation.)

In Fig.~\ref{fig3}c, under illumination at $\omega = 520$ nm (the Fano-enhancement wavelength), the hotspot becomes highly localized—about 1 nm wide—and significantly stronger beneath the QO.
This extreme confinement arises from the off-resonant Fano-enhancement effect, effectively restoring the near-field imaging resolution to the single-molecule scale. We utilize this effect in Fig.~\ref{fig4} where we scan two fluorescent nanoparticles .

\textit{Fano enhancement at off-resonance.}— We now show that the enhancement peak observed at $\omega = \omega_{\rm enh} = 520$ nm (2.384 eV) in Fig.~\ref{fig2}b—whose magnitude is comparable to the natural $\Omega_p\equiv\Omega_2=$597 nm (2.077 eV) peak—indeed originates from the cancellation of the off-resonant term $i(\Omega_p - \omega)$ in the denominator of Eq.~(\ref{FR}). We first demonstrate that the imaginary part of the QO-induced term enters with the opposite sign of $i(\Omega_p - \omega)$. We next support this interpretation with additional simulations.

The denominator of Eq.~(\ref{FR}) can be written as
\begin{equation}
	[i(\Omega_p-\omega)+\gamma_p] + i \, y|f|^2 / (\Omega_{\rm \scriptscriptstyle QO}-\omega),
	\label{FR_denominator}
\end{equation}
because $\gamma_{\rm \scriptscriptstyle QO}/\omega \sim 10^{-5}$ is negligibly small.
In our system the detunings are $(\Omega_p-\omega)=-0.307$ eV and $(\Omega_{\rm \scriptscriptstyle QO}-\omega)=-0.024$ eV. From previous studies, the Fano interference relevant for our system appears for a negative population difference $y = \rho_{ee}-\rho_{gg} < 0$~\cite{tacsgin2018fano,singh2016enhancement,postaci2018silent,wu2010quantum}.

Since $y<0$ and $(\Omega_{\rm \scriptscriptstyle QO}-\omega)<0$, the ratio $y|f|^2/(\Omega_{\rm \scriptscriptstyle QO}-\omega)$ is positive, and therefore the term $i\,y|f|^2/(\Omega_{\rm \scriptscriptstyle QO}-\omega)$ has the opposite sign of the off-resonant plasmon term $i(\Omega_p - \omega) = i(-0.307)$ eV.
Thus, the QO-induced term can indeed compensate the off-resonance shift.

Magnitude of the cancellation: The small detuning $|\Omega_{\rm \scriptscriptstyle QO} - \omega| = 0.024$ eV allows the QO term to become large enough to match the magnitude of the $-0.307$ eV plasmon detuning. The required cancellation takes place for a representative coupling strength $f \simeq 0.04,\omega$~\cite{wu2010quantum,leng2018strong,shah2013ultrafast}, where one obtains $i\, y \, |f|^2 / (\Omega_{\rm \scriptscriptstyle QO}-\omega)\approx$+$i\, 0.307$ eV thereby canceling the off-resonant contribution $i(\Omega_p - \omega)=-i\, 0.307$ in Eq.~(\ref{FR_denominator}).
As a result, the plasmonic mode with natural resonance at $\Omega_p\equiv\Omega_2=$597 nm is effectively driven on resonance at the off-resonant illumination wavelength $\omega = \omega_{\rm enh} = 520$ nm. This yields an intensity enhancement of $|E|/|E_0|\sim 170$ which matches the strength of the original $\Omega_p\equiv\Omega_2=$597 nm peak in Fig.~\ref{fig2}a.~\footnote{Note that Eq.~(\ref{FR_denominator}) also explains why the original 579 nm peak remains unaffected by the interference, consistent with Fig.~\ref{fig2}b. For illumination around $\omega \simeq 597$ nm, the detuning $(\Omega_{\rm \scriptscriptstyle QO}-\omega)$ becomes large, which suppresses the second term in the denominator and makes its contribution negligible at resonance.}

Thus, the 597 nm resonance is made to operate at 520 nm purely via Fano interference.
Such an off-resonant Fano enhancement occurs only directly beneath the QO. Other regions of the apex operate at the weaker near-field intensity corresponding to 520 nm in Fig.~\ref{fig2}a. See the near-field profile in Fig.~\ref{fig3}c. Thus, the scanning tip performs single-molecule-resolution a-SNOM as observed in Fig.~\ref{fig4}.

Verification with additional simulations:
We further vary $\Omega_{\rm \scriptscriptstyle QO}$ in separate simulations and track the positions of the resulting $\omega_{\rm enh}$ peaks. In all cases, the enhancement peaks have comparable height to the 597 nm resonance and arise from the same cancellation mechanism for similar values of $y|f|^2$~\cite{PSRetardation}: the imaginary part of the QO-induced term compensates the off-resonant term $i(\Omega_p - \omega)$ in the denominator.

These results confirm that the enhancement peak at 520 nm in Fig.~\ref{fig2}b originates from this cancellation, which effectively makes the $\Omega_2 = 597$ nm plasmon resonance operate as if it were on resonance at $\omega_{\rm enh} = 520$ nm.

%%%%%%%%%%%%%%%%%%%%%%%%%%%%%%%%%%%%%%%%%%%%%%%%%%%%%%%%%%%%%%%%%%%%%%%%%%%%%%%%%%%%%%%%%%%%%%%%%%%%%%%%%%%%%%%%%%%%%%%%%%%%%%%%%

\textit{In summary}, we present an experimentally feasible scheme for achieving single-molecule–resolution optical imaging under continuous-wave (CW) illumination. The setup consists of a gold-coated AFM tip covered with a 2D material. A stress-induced defect naturally forms at the point of maximum bending—the apex—and acts as a quantum object (QO). Similar configurations have already been demonstrated experimentally in Refs.~\cite{palacios2017large,branny2017deterministic,BendingDefectAPL2019,darlington2020imaging}.

Off-resonant Fano interference (enhancement) occurs only beneath the QO. At this location, the interference forces a much stronger plasmonic resonance (originally at $\Omega_2 = 597$ nm) to operate effectively at 520 nm (Fig.~\ref{fig2}b). In contrast, the remaining parts of the apex stay off-resonant with respect to the 597 nm mode under 520 nm excitation and therefore generate weaker near-field intensity. As a result, the field amplitude directly beneath the QO is about six times larger than that at the rest of the apex (Fig.~\ref{fig3}c), enabling single-molecule–resolution a-SNOM imaging, as demonstrated in Fig.~\ref{fig4}.

%Off-resonant Fano interference (enhancement) occurs only beneath the QO. At this location, the interference makes a much stronger plasmonic resonance (originally at $\Omega_2=$597 nm) effectively operate at 520 nm (Fig.~\ref{fig2}b). In contrast, other parts of the apex remain off-resonant to the stronger 597 nm mode at the 520 nm drive and therefore produce weaker near-field intensity. The field amplitude directly below the QO is about six times larger than at the rest of the apex (Fig.~\ref{fig3}c). As a result, single-molecule-resolution a-SNOM imaging becomes possible, as shown in Fig.~\ref{fig4}.

We show that this effective resonant operation at an off-resonant wavelength arises from a cancellation mechanism in the denominator of the plasmon amplitude: the QO-induced term compensates the off-resonant contribution between 597 nm and 520 nm (see Eq.~(\ref{FR_denominator}) and the related discussion).

The mechanism used here is fully compatible with standard a-SNOM operation under CW illumination. It is conceptually and practically distinct from lifetime-enhancement–based scheme~\cite{ovali2021single}, where ultrashort pulses introduce complications such as pulse-to-pulse interference and lingering plasmon excitations, making them far from conventional a-SNOM practice.

A further advantage of the present scheme lies in the tip geometry requirements. Achieving 10-nm-resolution near-field imaging conventionally demands ultra-sharp, gold-coated AFM tips fabricated with advanced nanolithography~\cite{alpan1,bazylewski2017review}. In contrast, our method remains effective even with larger apex sizes, since the 2D material–based defect formation technique demonstrated by Atatüre and co-workers~\cite{palacios2017large,branny2017deterministic,BendingDefectAPL2019,darlington2020imaging} naturally produces the QO at the tip apex, regardless of the overall tip width.

{\it Acknowledgments.}--- RVO and MET is supported by TUBITAK Grant No: 1001-121F141.

\bibliography{bibliography}
% Produces the bibliography via BibTeX.

\end{document}